\documentclass[sigconf]{acmart}
\AtBeginDocument{%
  \providecommand\BibTeX{{%
    \normalfont B\kern-0.5em{\scshape i\kern-0.25em b}\kern-0.8em\TeX}}}

\copyrightyear{2023}
\acmYear{2023}
\setcopyright{acmlicensed}
\acmConference[KDD '23]{Proceedings of the 29th ACM SIGKDD Conference on Knowledge Discovery and Data Mining}{August 6--10, 2023}{Long Beach, CA, USA}
\acmBooktitle{Proceedings of the 29th ACM SIGKDD Conference on Knowledge Discovery and Data Mining (KDD '23), August 6--10, 2023, Long Beach, CA, USA}
\acmPrice{15.00}
\acmDOI{10.1145/3580305.3599834}
\acmISBN{979-8-4007-0103-0/23/08}

\usepackage{natbib}
\usepackage{graphicx}
\usepackage{xcolor}
\usepackage[math]{alp}
\usepackage{ulem}
\usepackage{enumitem}
\usepackage{paralist}
\usepackage{algorithm}
\usepackage{algorithmic}
\usepackage{balance} 
\begin{document}

\newcommand{\xc}[1]{\noindent{\color{magenta} [Xiaohui: #1]}}
\newcommand{\az}[1]{\noindent{\color{blue} [Aonan: #1]}}
\newcommand{\jk}[1]{\noindent{\color{sky}[Jiankai: #1]}}
\newcommand{\rg}[1]{\noindent{\color{orange}[Ruocheng: #1]}}
\newcommand{\tw}[1]{\noindent{\color{red} [Taiqing: #1]}}
\newcommand{\lp}[1]{\noindent{\color{teal} [Liping: #1]}}
\newcommand{\ed}[1]{\noindent{\color{black} #1}}
\title{Graph-Based Model-Agnostic Data Subsampling for Recommendation Systems}

% \affiliation{%

% \affiliation{%

% \affiliation{%

\author{Xiaohui Chen}
\authornote{The majority of this work was done while these authors were at ByteDance.}
\affiliation{%
 \institution{Tufts University}
 \city{Medford}
 \state{Massachusetts}
 \country{USA}}
 \email{xiaohui.chen@tufts.edu}

\author{Jiankai Sun}
\affiliation{%
  \institution{ByteDance Inc.}
  % \streetaddress{30 Shuangqing Rd}
  \city{Bellevue}
  \state{Washington}
  \country{USA}}
  \email{jiankai.sun@bytedance.com}

\author{Taiqing Wang}
\affiliation{%
  \institution{ByteDance Inc.}
  % \streetaddress{8600 Datapoint Drive}
  % \streetaddress{30 Shuangqing Rd}
  \city{Bellevue}
  \state{Washington}
  \country{USA}}
\email{taiqing.wang@bytedance.com}

\author{Ruocheng Guo}
\affiliation{%
  \institution{ByteDance Research}
  % \streetaddress{8600 Datapoint Drive}
  % \city{San Antonio}
  \city{London}
  \country{UK}}
\email{ruocheng.guo@bytedance.com}

\author{Li-Ping Liu}
\affiliation{%
 \institution{Tufts University}
 \city{Medford}
 \state{Massachusetts}
 \country{USA}}
\email{liping.liu@tufts.edu}

\author{Aonan Zhang}
\authornotemark[1]
\affiliation{%
  \institution{Apple Inc.}
  \city{Seattle}
  \state{Washington}
  \country{USA}}
\email{aonan_zhang@apple.com}
\renewcommand{\shortauthors}{Xiaohui Chen et al.}



\begin{abstract}

    Data subsampling is widely used to speed up the training of large-scale recommendation systems. Most subsampling methods are model-based and often require a pre-trained pilot model to measure data importance via e.g. sample hardness. However, when the pilot model is misspecified, model-based subsampling methods deteriorate. Since model misspecification is persistent in real recommendation systems, we instead propose model-agnostic data subsampling methods by only exploring input data structure represented by graphs. Specifically, we study the topology of the user-item graph to estimate the importance of each user-item interaction (an edge in the user-item graph) via graph conductance, followed by a propagation step on the network to smooth out the estimated importance value. 
    Since our proposed method is model-agnostic, we can marry the merits of both model-agnostic and model-based subsampling methods. Empirically, we show that combing the two consistently improves over any single method on the used datasets.
    Experimental results on KuaiRec and MIND datasets demonstrate that our proposed methods achieve superior results compared to baseline approaches.
\end{abstract}



% \keywords{datasets, neural networks, gaze detection, text tagging}



\maketitle

\section{Introduction}
\label{sec:intro}
Recommendation systems learn user preferences through user-item interactions such as user clicks. Usually, a user click is considered as a positive sample that indicates the user's interest in the clicked item while a no-click between a user-item pair is considered as a negative sample. A click-through rate (CTR) prediction model outputs click probabilities of user-item pairs, and such probabilities can be used to rank items upon a user's request. A CTR model is trained using data collected from online platforms, where user-item pairs with no-clicks dominate. The imbalance of the dataset~\citep{chawla2009data} justifies  \textit{negative sampling} (NS), which down-samples negative samples and significantly reduces model training costs.

% \xc{Collaborative filtering (CF) in recommender systems aims to learn user preferences through user-item interactions. Nowadays, online platforms collect enormous amounts of user-item interaction every day, and only a small portion of the data indicates the users' interests, i.e., the positive samples. The imbalance of the dataset justifies the use of negative sampling (NS), which can significantly reduce the cost of training a model. }

% \az{Recommendation systems learn user preferences through user-item interactions such as user clicks. Usually, user click signals are identified as positive samples that indicate user's strong interest toward items and no-click signals are negative samples. A clickthrough rate (CTR) prediction model output the probability of a user-item click which can be utilized to rank items given a user's request. Since no-clicks dominates, the imbalance of the dataset~\citep{chawla2009data} justifies the use of negative sampling (NS), which significantly reduce model training cost.}

\begin{figure*}
    \centering
    \includegraphics[width=\textwidth]{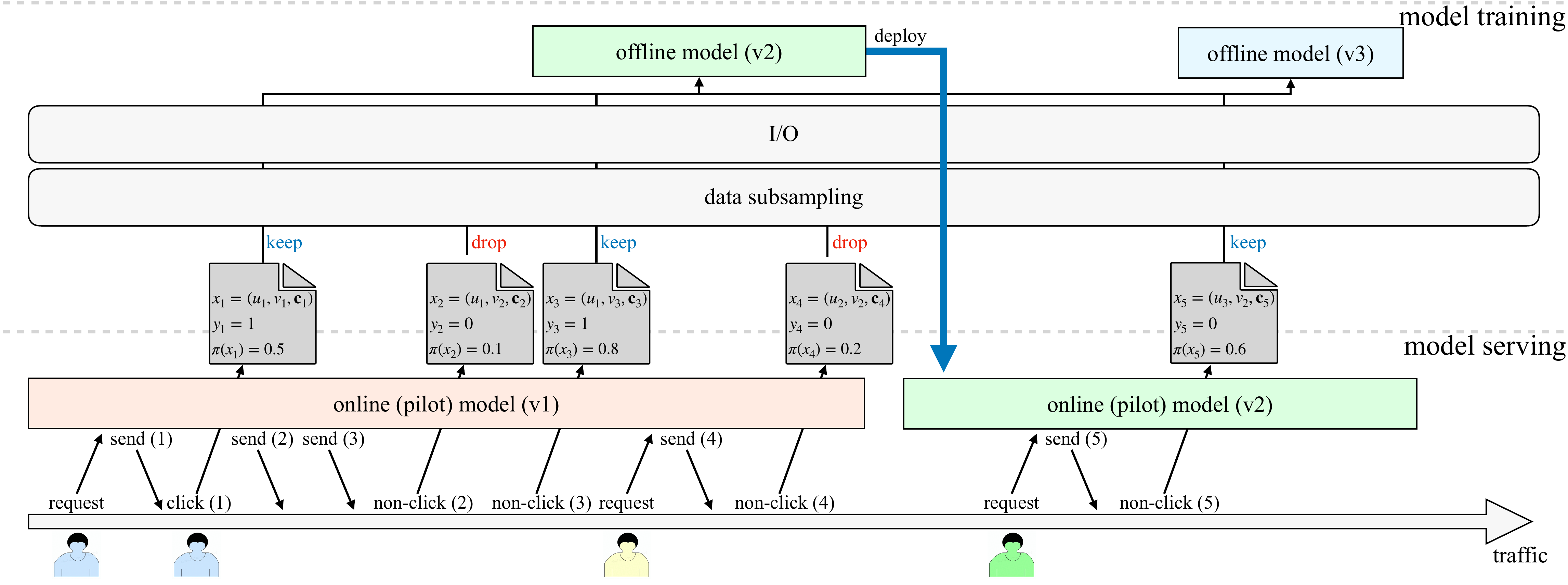}\vspace{5pt}
    \caption{A simplified recommendation system. A user initiates a request to the online serving model and receives recommendations returned by the server. User clicks will trigger positive instances, while non-clicks will trigger negative instances. All instances are filtered by the data subsampling module. For model-based methods, subsampling rates are determined by online models (e.g. model v1), thus sub-optimal for offline training (e.g. model v2) purposes. When model v2 is deployed online, data subsampling is affected, producing inconsistent subsampling rates.}
    \label{fig:deploy_demo}
\vspace{-0em}
\end{figure*}

In contrast to treating all data points as equally important, non-uniform subsampling aims to obtain more informative samples. Previous model-based methods~\citep{zhang2013optimizing, fithian2014local, ting2018optimal, han2020local, ding2020simplify, wang2020logistic, wang2021nonuniform} utilize a pilot model to assess the importance of data. When the pilot model is correctly pre-trained, one can achieve the optimal sampling rate: the method is measuring the importance of data using pilot prediction scores together with first and second-order derivatives of the loss function~\citep{wang2021nonuniform}. Since optimal negative sampling rates are proportional to the pilot model's prediction score, a high sampling rate indicates an inaccurate model prediction. Thus, one can interpret the sampling strategy as using hard negative samples (HNS)~\citep{zhang2013optimizing,ding2020simplify}.

We find that model-based sampling algorithms may not be applicable in real industrial scenarios. Figure~\ref{fig:deploy_demo} demonstrates how a real recommendation system deploys data subsampling. A user initiates a request to the online serving model and receives recommendations returned by the server. If the user clicks the recommended item, then a positive instance is collected. Otherwise, if the user does not click the item within a period of time, a negative instance is collected. Note that the pilot prediction score and other statistics are recorded in the instance to calculate the sampling rate.
All instances are filtered by the data subsampling module before I/O to reduce the I/O and network bandwidth bottleneck. Offline models are trained with historical data before being deployed online.

In these scenarios, there are two unavoidable obstacles to the application of model-based sampling. First, offline model training is vulnerable to model misspecification, which leads to inferior results. Unfortunately, model misspecification is persistent due to online-offline discrepancy, especially in continuous integration and deployment (CI/CD)~\citep{shahin2017continuous} in real systems. Second, coupling data subsampling and model training introduces extra dependencies across system modules, which increases system maintenance cost and brings about extra technical debt~\citep{sculley2015hidden}.

To maintain a scalable and sustainable data subsampling service, we propose model-agnostic data subsampling methods based on user-item bipartite graphs. In the bipartite graph, two sets of nodes represent users and items separately, and each edge represent interactions between a user and an item. Then we treat edges as instances to consider subsampling. In the context of a bipartite graph, we reinterpret the idea of HNS using effective conductance~\citep{chung1997spectral}. An edge is considered a hard instance if the effective conductance over its two nodes is high. With this notion of hardness, we assign high sampling rates to edges with high effective conductance.
Additionally, we exploit the bipartite graph to propagate and refine the sampling rates. Since our proposed method is model-agnostic, we can marry the merits of both model-agnostic and model-based subsampling methods. Empirically, we show that combing the two consistently improves over any single method on the widely used datasets. To summarize our contribution:
\begin{compactitem}
    \item we propose a model-agnostic method to measure data hardness scores via effective conductance on the user-item bipartite graph;
    % \item we construct user-item bipartite graphs from the data, where edges represent instances; \az{Is this a contribution?}
    % \item we propose to use effective conductance on graphs to measure data hardness scores (sampling rates);
    \item we refine hardness scores via propagation on graphs;
    \item empirical results demonstrate that without requesting information from the pilot model, our model-agnostic methods achieve comparable or better results than model-based methods;
    \item we verify model-agnostic methods compliment model-based methods and combining them can achieve even better performance. %we verify 
\end{compactitem}

\section{Preliminaries}
\label{sec:preliminaries}
We consider a binary classification problem and let $\calD=\{(\bx_n, y_n)\}_{n=1}^N$ be a training set of size $N$. Here $\bx_n$ and $y_n$ are, respectively, the feature vector and label of instance $n$. %$\bx_n$ represent features, and $y_n\in\{0, 1\}$ represents the user's feedback (e.g. click or not) on the recommended item. 
We study the generalized logistic regression (GLM) model, where the target model corresponding to the offline model (e.g., model v2 before deployment in Figure~\ref{fig:deploy_demo}) is represented as
\begin{align}
    f(\bx;\theta) := p(y=1|\bx,\theta) = \frac{1}{1+e^{-g(\bx;\theta)}},
\end{align}
where the log-odd $g(\bx;\theta)$ is implemented by a predictive model. Denote $N_0$ as the number of negative instances and $N_1=N-N_0$ as the number of positive instances. We study the case where $\calD$ is imbalanced, i.e., $N_0\gg N_1$. Since information is sparsely distributed over many negative instances~\citep{wang2020logistic}, we often use NS to reduce the dataset size and boost training efficiency.

\begin{algorithm}[t]
  \caption{Hard Negative Sampling (HNS)}\label{alg:hardness}
\begin{algorithmic}
\STATE Given $\{(\bx_n,y_n)\}_{n=1}^N$ and hardness score function $h(\cdot)$
\STATE Compute the counterfactual NSR $\pi(\bx_n)$ via Eqn.~(\ref{alg:sample_rate}).
\FOR {$n=1, ..., N$}
\STATE Generate $\lambda_n\sim \mathbb{U}(0,1)$.
\IF {$y_n=1$ or $\lambda_n\le\pi(\bx_n)$} 
\STATE Include $\{\bx_n,y_n, \pi(\bx_n)\}$ in the training set.
\ENDIF
\ENDFOR
\end{algorithmic}
\end{algorithm}

\begin{figure*}
    \centering
    \includegraphics[width=\textwidth]{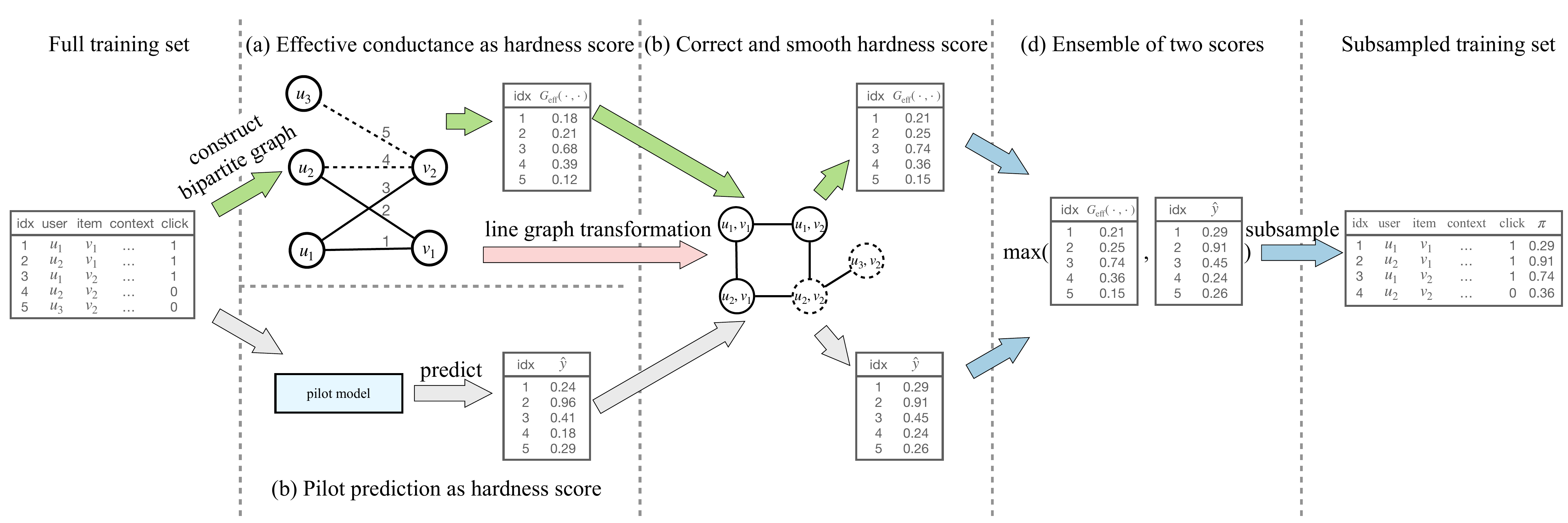}\vspace{5pt}
    \caption{Illustration of our model-agnostic subsampling framework. We construct the user-item bipartite graph from the training data and estimate the sample hardness using effective conductance (a); The user-item graph can be transformed into a line graph, and smooth the hardness score through propagation (c); Similarly, we can smooth the hardness score obtained from a model-based method (b); The smooth scores (a-c,b-c) can be further ensembled to calculate the final subsampling rate (d).}
    % In the user-item graph, sample (edge) hardness can be estimated using effective conductance (a); the hardness scores can be further smoothed via edge propagation (c);  }
    \label{fig:work-flow}
\vspace{-0em}
\end{figure*}
An NS algorithm weighs each negative instance $\bx$ with some measurement of its importance $\pi(\bx)$. The importance $\pi(\bx)$ is used as the negative sampling rate (NSR) of $\bx$. The assignment of $\pi(\bx)$ follows a widely-used heuristic, which is exploiting  ``hard negative samples''~\citep{robinson2020contrastive,huang2020embedding}. Sampling rates are proportional to non-negative hardness scores $h(\cdot)$:
\begin{align}
    \pi(\bx_n) \propto h(\bx_n),\quad \text{s.t.}~\sum_{n=1}^{N_0}\pi(\bx_n) = N_0\alpha,
    \label{alg:sample_rate}
\end{align}
where $\alpha\in(0,1]$ is a pre-set average subsampling rate of negative instances. Suppose we have a pilot model $\tilde{f}(\cdot)$ corresponding to online model v1 in Figure ~\eqref{alg:sample_rate}. When $\tilde{f}(\cdot)=f(\cdot; \theta^*)$, that is $\tilde{f}(\cdot)$ has the same functional form as the target model, and $\theta^*$ is the true parameter. Then we can set a \textit{model-based hardness score} $h_b(\bx_n) = \tilde{f}(\bx_n) = f(\bx_n; \theta^*)$  to get a near-optimal sampling rate $\pi(\bx_n)$ by Eqn.~\eqref{alg:sample_rate}~\citep{wang2021nonuniform}. Intuitively, a negative instance predicted with a higher score by the pilot model $\tilde{f}(\cdot)$ is more ``surprising'' and thus is harder for the target model $f(\cdot;\theta)$. Note that we use $\pi(\bx)$ for a positive instance to denote its counterfactual negative sampling rate. We demonstrate the HNS procedure in Algorithm~\ref{alg:hardness}. 

Since data distribution shifts after subsampling, one needs to correct the log odds to get an unbiased estimation:
\begin{align}\hspace{-2pt}
    \hat{\theta} = \arg\max_\theta \sum_{n=1}^N \delta_n[y_n g(\bx_n;\theta)-\log(1+e^{g(\bx_n;\theta)-\ell_n)}],
\label{eq:logits_correction}
\end{align}
where $\delta_n\in\{0,1\}$ is the subsampling indicator and $\ell_n := \log\pi(\bx_n)$. \cite{han2020local} proves that log odds correction is more efficient than the IPW estimator~\citep{horvitz1952generalization}.

However, when the pilot model is misspecified, optimal NS with pilot models is not achieved. In real scenarios, it may be error-prone to deploy model-based HNS methods
% \tw{this statement might be challenged by reviewers with industry background. model-based approaches are used in industry indeed. I would say it may be error-prone / vulnerable to deploy model-based methods. this statement also invalidates our approach to combine model-based and model-agnostic approaches.} \az{sure will change later.} 
since we persistently suffer from model misspecification problems due to online-offline model discrepancy and continuous integration and deployment. Thus, it is tempting to use \textit{model-agnostic hardness score} $h_a(\cdot)$ to maintain a scalable and sustainable data subsampling service.

\section{Methodology}
\label{sec:methodology}
\vspace{-0.0em}

An overview of our workflow is demonstrated in Figure ~\ref{fig:work-flow}. 
We introduce its technical details step by step. 

\vspace{5pt}
\subsection{MA-EC: \textit{M}odel-\textit{a}gnostic Negative Hardness Estimation via \textit{E}ffective \textit{C}onductance}
\label{sec:agnostic}\vspace{-0em}
In this section, we consider the subsampling problem in the context of a bipartite graph formed from a recommendation problem.
Let the bipartite graph be $(U, V, E)$, where the node set $U=\{u_i\}_{i=1}^M$ represents $M$ users, the node set $V=\{v_j\}_{j=1}^Q$ represents $Q$ items, and the edge set $E=\{(u_{i_n},v_{j_n})\}_{n=1}^N$ represent $N$ user-item pairs. For each node pair $n$, $y_n\in\{0,1\}$ represents whether there is a positive interaction between $u_{i_n}$ and $v_{j_n}$.

In our classification problem, each feature $\bx_n = (u_{i_n},v_{j_n},\bc_n)$, where $\bc_n$ represent context features. And its label is $y_n$. We aim to compute the model-agnostic hardness $h_a(\cdot)$ of negative samples without referring to a pilot model.
% \jk{should be $M$ and $N$? user and item do not have the same number?\xc{it's a union of user/item nodes over all N instances, the index n means the index of instance, not the index of node}}

% This section elaborates on how to evaluate the hardness of negative samples by exploring the training set $\calD$. Instead of using a pilot model, we define the hardness scoring function $h(\cdot):=f(\cdot;\calD)$ \rg{in a data-centric manner}. Specifically, denote $G=(V, E)$ to be the bipartite user-item graph over the dataset $\calD$, where $V=\{v_i\}_{i=1}^{|V|}=\{u(n)\}\cup\{o(n)\}$ are the nodes including all $N_u$ users and $N_o$ items. $E=\{e_{(u(n),o(n))}\}_{n=1}^N$ are the edges, in which each edge \rg{$e_{(u(n),o(n))}$} corresponds to the interaction $\bx_n=(u(n),o(n),\bc_n)$. In the following, the notation of $n$ and $(u(n),o(n))$ will be used interchangeably.

% \az{This section elaborates on evaluating the hardness of negative samples without referring to a pilot model. Instead, we define the hardness score function in a data-centric manner as $h(\cdot):=f(\cdot;\calD)$. Specifically, we simplify data structure as a user-item bipartite graph $(U, V, E)$, where $U=\cup_{n=1}^N\{u_n\}$, $V=\cup_{n=1}^N\{v_n\}$ represent user and item nodes separately. $E=\{e_{(u_n,v_n)}\}_{n=1}^N$ are the edges, in which each edge $e_{(u_n,v_n)}$ corresponds to the instance $\bx_n=(u_n,v_n,\bc_n)$. In the following, we use $e_n$ as a shorthand for $e_{(u_n,v_n)}$.}

\begin{figure}[t]
    \centering
    \includegraphics[width=0.45\textwidth]{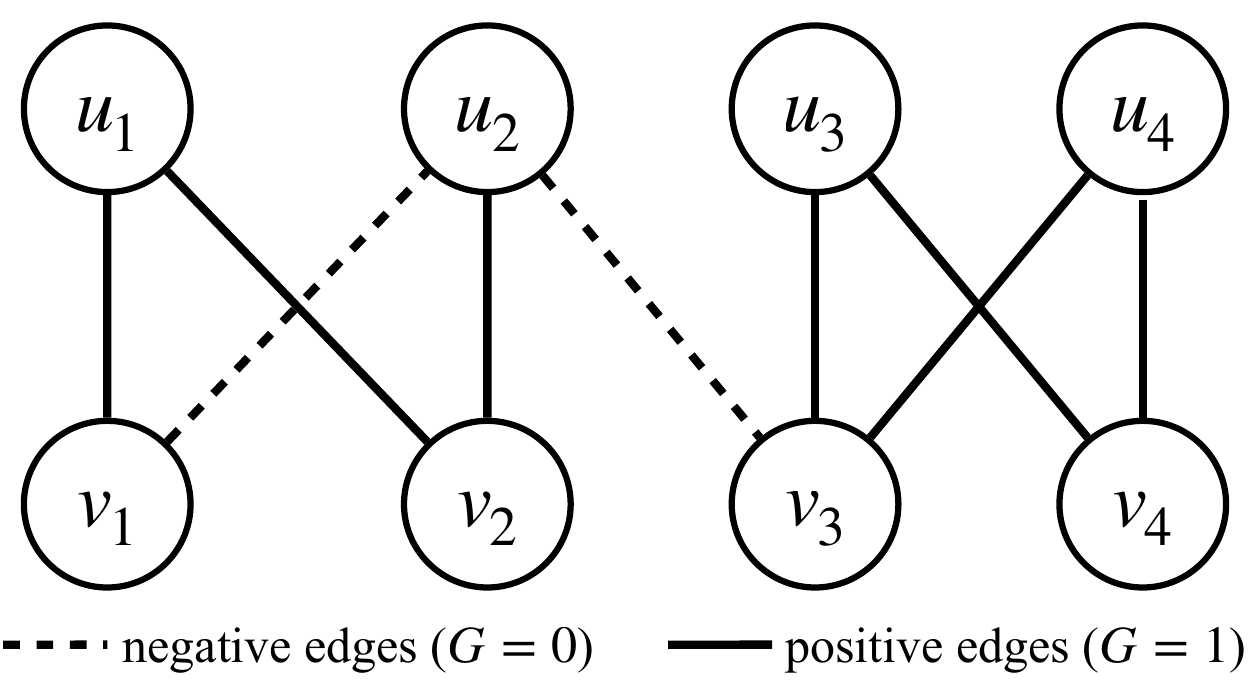}\vspace{-0pt}
    \caption{Interpret effective conductance on a bipartite graph. All positive edges are assigned conductances $G\!\!=\!\!1$, and all negative edges have $G\!\!=\!\!0$. Consider two user-item pairs $(u_2,v_1),(u_2,v_3)$ and their effective conductances $G_\mathrm{eff}(u_2,v_1)\!\!=\!\!1/3$, $G_\mathrm{eff}(u_2,v_3)\!\!=\!\!0$. Effective conductances demonstrate user preference. We observe a 3-hop path between $u_2$ and $v_1$ $(u_2\!-\!v_2\!-\!u_1\!-\!v_1)$, but no path between $u_2$ and $v_3$, thus $u_2$ may prefer $v_3$ over $v_1$. $(u_2,v_1)$ corresponds to a harder negative sample than $(u_2,v_3)$.}
    \label{fig:conductance-resistance}
\vspace{-0em}
\end{figure}

We relate sample hardness to graph topology. Imagine the user-item graph as an electricity network, where each edge $(u_{i_n}, v_{j_n})$ is a conductor with conductance $G(u_{i_n},v_{j_n})$, which measures the \textit{edge}'s ability in transferring electrical current. We expect $G(u_{i_n},v_{j_n})$ to be large when user $u_{i_n}$ expresses direct preference of item $v_{j_n}$. Particularly, we set 
\begin{align}
G(u_{i_n},v_{j_n}) = y_n.
\end{align}
It means that the conductance is one if there is a direct preference or 0 otherwise. 
% We can use other numbers to express relative conductance values for $y_n=1$ and $y_n = 0$, but here we just use $y_n$ for simplicity.\xc{I think this sentence can be removed}

Then we consider the \textit{effective conductance} $G_\mathrm{eff}(u_{i_n}, v_{j_n})$ between $u_{i_n}$ and $v_{j_n}$. It represents the \textit{network}'s ability to transfer current from $u_{i_n}$ to $v_{j_n}$  (or the opposite direction). The effective conductance $G_\mathrm{eff}(u_{i_n}, v_{j_n})$ is the reciprocal of effective resistance $R_\mathrm{eff}(u_{i_n}, v_{j_n})$: they are defined as follows~\citep{alexander2007fundamentals}.
\begin{align}
R_\mathrm{eff}(u_{i_n}, v_{j_n}) &\!=\! (\be[u_{i_n}] \!-\! \be[v_{j_n}])^{\top} \bL^{+} (\be[u_{i_n}] - \be[v_{j_n}]), \nonumber\\
G_\mathrm{eff}(u_{i_n}, v_{j_n}) &\!=\! \frac{1}{R_\mathrm{eff}(u_{i_n}, v_{j_n})}.
\end{align}
Here $\be[\cdot] \in \{0, 1\}^{M+Q}$ is the one-hot encoding of a node in the graph, and $\bL^{+}$ is the pseudo inverse of the Laplacian of the graph. Intuitively, if there are many conductible paths between $u_{i_n}$ and $v_{j_n}$, then the effective conductance $G_\mathrm{eff}(u_{i_n}, v_{j_n})$ is large.

\paragraph{Estimating sample hardness via effective conductance.}
\iffalse
We use the difference between effective conductance and conductance of an edge to decide the hardness score of the instance.  
For a negative sample $(\bx_n,y_n=0)$, the hardness score as\xc{for positive samples this rule also applies}
\begin{align}
    h_a(\bx_n) = G_\mathrm{eff}(u_{i_n},v_{j_n})-G(u_{i_n},v_{j_n}).
    \label{eq:conductance_hardness}
\end{align}
Note that $h_a(\bx_n)$ is always positive because the effective conductance is always larger than the conductance between the same  pair of nodes.

The difference $h_a(\bx_n)$ indicates the hardness of negative instances for the following reason. Suppose $y_n = 0$ and there is no interaction between $u_{i_n}$ and $v_{j_n}$, then $G(u_{i_n},v_{j_n}) = 0$. However, if $G_\mathrm{eff}(u_{i_n},v_{j_n})$ is larger, then it means  there are multiple high-conductance paths from $u_{i_n}$ to $v_{j_n}$. For example, $u_{i'}$ clicks many items including $v_{j_n}$, and $u_{i_n}$ clicks many other items clicked by $u_{i'}$ but not $v_{j_n}$. In these cases, the lack of interaction ($y_n = 0$) between $u_{i_n}$ and $v_{j_n}$ is unexpected, and thus this instance is a hard one for the classifier.  Figure~\ref{fig:conductance-resistance} demonstrates an example.
\fi

Figure~\ref{fig:conductance-resistance} demonstrates that effective conductance positively relates to sample hardness. Define the hardness score as
%\xc{for positive samples this rule also applies}
\begin{align}
    h_a(\bx_n) = G_\mathrm{eff}(u_{i_n},v_{j_n})-G(u_{i_n},v_{j_n}).
    \label{eq:conductance_hardness}
\end{align}
For a negative sample, $G(u_{i_n},v_{j_n})=0$. The effective conductance $G_\mathrm{eff}(u_{i_n},v_{j_n})$ is high when there are multiple high-conductance paths from $u_{i_n}$ to $v_{j_n}$, demonstrating user’s indirect preference to the item. When the indirect preference is high but $(u_{i_n},v_{j_n})$ turns out to be negative, we identify it as a hard negative sample. For a positive sample, $h_a(\bx_n)$ denotes its counterfactual hardness score by subtracting the direct conductor $G(u_{i_n},v_{j_n})$ from $G_\mathrm{eff}(u_{i_n},v_{j_n})$ to eliminate the prior information given by the label. The hardness score is used to calculate the counterfactual NSR for log odds correction in Eqn.~(\ref{eq:logits_correction}). We do not drop positive samples.

\paragraph{Implementation.} Direct calculation of effective conductance is time-consuming. Instead, we first approximate the commute time distance $\mathrm{comm}(u,v)$ through random walk using scientific computing tools~\citep{staudt2016networkit}. Then we use the transformation $G_{\mathrm{eff}}(u,v)=2|E|/\mathrm{comm}(u,v)$~\citep{chandra1996electrical} to convert the commute time into effective conductance.
\subsection{Smoothing Hardness Scores through Edge Propagation}
\label{sec:propagation}\vspace{-0em}
% Since we set $w(y=0)=0$, the graph to calculate effective conductance only contains positive edges and is highly sparsified. According to Eqn.~(\ref{eq:conductance_hardness}), any isolated edges will have zero hardness score.
% This section shows this problem can be alleviated by propagating edge hardness scores over the entire graph with positive and negative edges.
% When estimating the sample hardness of user-item pairs via effective conductance, some pairs might have zero effective conductance and get meaningless subsampling rates. For example, edges connected to users who negatively respond to all recommended items; or edges connected to items that receive no positive responses. This section shows that this problem can be alleviated by propagating edge hardness scores over the graph.

% \az{Calculating effective conductance with positive edges will sparsify the graph, resulting in isolated edges with zero hardness score according to Eqn.~(\ref{eq:conductance_hardness}). This section shows this problem can be alleviated by propagating edge hardness scores over the entire graph with positive and negative edges.}
The effective conductance derived from the sparsified graph is noisy, leading to an inaccurate estimation of hardness scores\footnote{Since negative edges are ignored to calculate effective conductance. According to Eqn.~(\ref{eq:conductance_hardness}), any isolated positive edges without a multi-hop connection between its endpoints will have zero hardness score without graph propagation.}. We use the idea of graph propagation to smooth the hardness score. While existing machine learning literature focuses on propagating node attributes (e.g., node features or labels) to smooth out node-level uncertainty~\citep{jia2020residual, huang2020combining}, propagating edge attributes is underexplored. 

We reduce edge propagation to node propagation by transforming user-item bipartite graph $(U,V,E)$ into its corresponding line graph $L(U,V,E)=(V_L,E_L)$, where $V_L=E$ and $E_L$ is the collection of edge pairs that share the same node~\citep{harary1960some}. 

\paragraph{Uncertainty propagation.} 
Denote $G_{\mathrm{eff}}:=(G_{\mathrm{eff}}(u_{i_n},v_{j_n}))_{n=1}^N\in\mathbb{R}^N$, $ Y:=(y_n)_{n=1}^N\in\{0,1\}^N$ to be the vector of the effective conductance scores and edge labels, respectively.
Similar to~\cite{jia2020residual}, we normalize the effective conductance $G_{\mathrm{eff}}$ as the estimated score $Z$ and calculate the uncertainty score $B$ as the absolute residual between $Z$ and $Y$
\begin{align}
    Z= \frac{G_{\mathrm{eff}}-\mathrm{min}(G_{\mathrm{eff}})}{\mathrm{max}(G_{\mathrm{eff}})-\mathrm{min}(G_{\mathrm{eff}})}, \quad B=|Y-Z|.
    \label{eq:uncertainty_score}
\end{align}
% The key idea here is the assumption that neighbouring \rg{neighboring (for American English)} nodes in a graph share similar uncertainty of the hardness estimation. Similar to~\cite{jia2020residual}, we use the absolute residual between the estimated score and the true label to quantify the uncertainty. To do so, we first need to normalize the effective conductance $C$ and obtain the uncertainty vector $B$
% \begin{align}
%     Z= \frac{C-\mathrm{min}(C)}{\mathrm{max}(C)-\mathrm{min}(C)}, \quad B=|Y-Z|.
% \end{align}
% \az{Assume neighboring nodes share similar uncertainty of the hardness estimation. Similar to~\cite{jia2020residual}, we use the absolute residual between the estimated score and the true label to quantify the uncertainty. We first normalize the effective conductance $G_{\mathrm{eff}}$ and obtain the uncertainty score $B$
% \begin{align}
%     Z= \frac{G_{\mathrm{eff}}-\mathrm{min}(G_{\mathrm{eff}})}{\mathrm{max}(G_{\mathrm{eff}})-\mathrm{min}(G_{\mathrm{eff}})}, \quad B=|Y-Z|.
%     \label{eq:uncertainty_score}
% \end{align}}
Here we use the min-max normalization to restrict the hardness score to be within the range $[0,1]$.  Denote $S=D_L^{-1/2}A_LD_L^{-1/2}$, where $A_L, D_L$ are the adjacency matrix and the degree matrix of the line graph, respectively. We smooth the uncertainty by solving the following optimization problem~\citep{zhou2003learning}:
\begin{align}
    \hat{B}=\underset{W}{\mathrm{argmin}}~\mathrm{tr}(W^T(I-S)W) + \mu||W-B||^2_F.
    \label{eq:uncertainty_opt}
\end{align}
% Here we use the min-max normalization to ensure that every hardness score is in $[0,1]$. Denote $S=D^{-1/2}AD^{-1/2}$ \rg{$S=D_L^{-1/2}A_LD_L^{-1/2}$}, where $A_L, D_L$ are the adjacency matrix and the degree matrix of the line graph, respectively. We smooth the uncertainty by directly solving the optimization problem~\citep{zhou2003learning}:
% \begin{align}
%     B^*=\underset{Q}{\mathrm{argmin}}~\mathrm{trace}(Q^T(I-S)Q) + \mu||Q-B||^2_F.
% \end{align}
% \az{Here we use the min-max normalization to restrict hardness score within the range $[0,1]$.  Denote $S=D_L^{-1/2}A_LD_L^{-1/2}$, where $A_L, D_L$ are the adjacency matrix and the degree matrix of the line graph, respectively. We smooth the uncertainty by directly solving the following optimization problem~\citep{zhou2003learning}:
% \begin{align}
%     \hat{B}=\underset{Q}{\mathrm{argmin}}~\mathrm{tr}(Q^T(I-S)Q) + \mu||Q-B||^2_F.
%     \label{eq:uncertainty_opt}
% \end{align}}
The first term in Eqn.~(\ref{eq:uncertainty_opt}) restricts the difference of uncertainties in neighboring nodes. And the second term constrains the smoothed uncertainty to be close to the initial uncertainty, with the coefficient $\mu$ controlling the strength of the constraint. With the smoothed uncertainty vector $\hat{B}$, we correct the hardness estimation by reversing  Eqn.~(\ref{eq:uncertainty_score})
\begin{align}
    \hat{Z} = Y+ (-1)^Y\hat{B}.
\end{align}
% The first term of the cost function means that uncertainties in neighboring nodes should not change too much. And the second term constrains the smoothed uncertainty to be close to the initial uncertainties, with the coefficient $\mu$ controlling the strength of the constraint. With the smoothed uncertainty vector $B^*$, we can correct the hardness estimation
% \begin{align}
%     \hat{Z} = Y+ (-1)^YB^*.
% \end{align}
% \az{The first term in Eqn.~(\ref{eq:uncertainty_opt}) restricts the difference of uncertainties in neighboring nodes. And the second term constrains the smoothed uncertainty to be close to the initial uncertainties, with the coefficient $\mu$ controlling the strength of the constraint. With the smoothed uncertainty vector $B^*$, we correct the hardness estimation by reversing  Eqn.~(\ref{eq:uncertainty_score})
% \begin{align}
%     \hat{Z} = Y+ (-1)^Y\hat{B}.
% \end{align}}
\cite{zhou2003learning,huang2020combining} also introduced an iterative approximation approach. Let $\gamma=1/(1+\mu)$ and
\begin{align}
    B^{t+1} = (1-\gamma) B + \gamma SB^{t},\quad B^0=B.
\end{align}
Then $B^t\rightarrow\hat{B}$ when $t\rightarrow\infty$. 
% \cite{zhou2003learning,huang2020combining} have shown that $B^*$ can be solved by performing the following iterations until convergence, let  $\gamma=1/(1+\mu)$, we have
% \begin{align}
%     B^{t+1} = (1-\gamma) B + \gamma SB^{t},\quad B^0=B.
% \end{align}

% \az{\cite{zhou2003learning,huang2020combining} also introduced an iterative approximation approach. Let $\gamma=1/(1+\mu)$ and
% \begin{align}
%     B^{t+1} = (1-\gamma) B + \gamma SB^{t},\quad B^0=B.
% \end{align}
% Then $B^t\rightarrow\hat{B}$ when $t\rightarrow\infty$. 
% }

However, this iterative approach is not scalable as the transformed line graph has $|E_L|=(\sum_{u_i\in U}\mathrm{Deg}(u_i)^2 + \sum_{v_j\in V}\mathrm{Deg}(v_j)^2)/2-N$ edges in total, where $\mathrm{Deg}(\cdot)$ represents node degree. Alternatively, we can directly propagate edge uncertainty along the original graph $(U,V,E)$, which only contains $|E|=(\sum_{u_i\in U}\mathrm{Deg}(u_i) + \sum_{v_j\in V}\mathrm{Deg}(v_j))/2$ edges in total. The propagation rule over edges is as follows
\begin{align}\hspace{-15pt}
    &B_n^{t+1} \!=\! (1-\gamma) B_n + \gamma\frac{m^t(u_{i_n}) + m^t(v_{j_n})-2Z_n}{\mathrm{Deg}(u_{i_n})+\mathrm{Deg}(v_{j_n})-2},\\
    &\text{where}~m^t(u) = \!\!\!\!\sum_{n:~u=u_{i_n}}\!\!\!\!B_n^t,~m^t(v) = \!\!\!\!\sum_{n:~v=v_{j_n}}\!\!\!\!B_n^t,~B_n^0=B_n.\nonumber
\end{align}
Using message passing mechanisms, we store the aggregated uncertainty $m^t(u)$ in $u$ and then update the uncertainty $B^{t+1}$ by applying the rule above.

\paragraph{Score propagation.} Instead of propagating uncertainty, we can directly propagate the scores $\hat{Z}$ by iterating
\begin{align}
    Z^{t+1} = (1-\gamma) \hat{Z} + \gamma SZ^{t},\quad Z^0=\hat{Z}
\end{align}
until convergence. After obtaining the final hardness scores, we rescale them to match the average subsampling rates $\alpha$. We find that smoothing hardness scores benefits both model-agnostic and model-based methods.

% We can further propagate the corrected scores $\hat{Z}$ on the graph directly. The assumption is similar to that of the uncertainty propagation -- we expect that nearby nodes to have similar prediction scores. We obtain the final smoothed scores $Z^*$ by iterating
% \begin{align}
%     Z^{t+1} = (1-\gamma) \hat{Z} + \gamma SZ^{t},\quad Z^0=\hat{Z}
% \end{align}
% until convergence. After obtaining the final hardness scores, we scale them to proper subsampling rates given the overall subsampling rate $\alpha$, similar to the model-based methods (which we will discuss in Section~\ref{}).

% Besides our model-agnostic method, we find that smoothing hardness scores also benefits the model-based method. We will further demonstrate it later in the experiments.

% \az{Instead of propagating uncertainty, we can directly propagate the scores $\hat{Z}$ by iterating
% \begin{align}
%     Z^{t+1} = (1-\gamma) \hat{Z} + \gamma SZ^{t},\quad Z^0=\hat{Z}
% \end{align}
% until convergence. After obtaining the final hardness scores, we rescale them to match the overall subsampling rates $\alpha$.

% We find that smoothing hardness scores benefits both model-agnostic and model-based methods. We will demonstrate empirical results in experiments.}
\vspace{-0em}
\subsection{Combining Model-agnostic and Model-based Methods}
\vspace{-0em}
\label{sec:ensemble}

% \jk{should not say empirically, we found it's effective. should say one is model-agonistic and one is MB. orthogonal. then combine them together. better not to show experimental results in the methodology. }

% As we will demonstrate in experiments, model-agnostic (MA) subsampling tends to select different portions of samples compared to model-based (MB) subsampling. Thus, it is tempting to ensemble MA and MB to further improve sample efficiency. 
% To further improve the performance, we investigate several strategies for combining the subsampling rate estimated from a model-based method and our model-agnostic method. We expect that the different methods score samples from different perspectives and thus complement each other. \rg{Maybe you want to mention this is verified by experimental results in xxx}

% \az{As we will demonstrate in experiments, model-agnostic (MA) subsampling tends to select different portions of samples compared to model-base (MB) subsampling. Thus, it is tempting to ensemble MA and MB to further improve sample efficiency.}

% Given the hardness score $h(\bx)$ for sample $\bx$, we set the least \rg{minimal?} sampling rate $\varrho$ and tune a linear scaling parameter $\rho$ to meet the overall subsampling rate $\alpha$
% \begin{align}
%     \pi(\bx) = \mathrm{max}(\mathrm{min}(\rho h(\bx),\varrho), 1),~ \text{s.t.}~\sum_{n=1}^{N_0}\pi(\bx_n) = N_0\alpha.
% \end{align}

We have described our model-agnostic data subsampling methods in previous sections. % Since both model-agnostic and model-based subsampling methods are orthogonal\xc{how to understand orthoganal}, 
Some hard instances may be overlooked by model-agnostic methods while they can be captured by model-based methods. Hence, in this section, we show how to combine these two methods to marry the merits of both and achieve a better sampling performance.

Given a sample $\bx$, model-agnostic and model-based subsampling methods calculate their corresponding sampling rate $\pi_\calD(\bx)$ and $\pi_{\phi}(\bx)$ respectively.  Particularly  $\pi_{\phi}(\bx)$ is  the subsampling rate for $\bx$ by using pilot model $h_b(\cdot):=\tilde{f}(\cdot;\phi)$, and $\pi_\calD(\bx)$ is the subsampling rate using model-agnostic hardness score $h_a(\cdot)$ in Eqn.~\eqref{eq:conductance_hardness}:
\begin{align}\hspace{-10pt}
\label{eq:pi_x}
    \pi_\phi(\bx) &\propto \mathrm{max}\Big(\mathrm{min}\big(\rho_\phi h_b(\bx),\varrho_\phi\big), 1\Big),\nonumber\\
    \pi_\calD(\bx) &\propto \mathrm{max}\Big(\mathrm{min}\big(\rho_\calD h_a(\bx),\varrho_\calD\big), 1\Big).
\end{align}
$(\varrho_\phi,\varrho_\calD)$ is the minimum sampling rate, and $(\rho_\phi,\rho_\calD)$ are tuned to meet the average subsampling rate $\alpha$.

% \begin{align}\hspace{-5pt}
% \label{eq:pi_x}
%     \pi(\bx) \!\!=\!\! \mathrm{max}(\mathrm{min}(\rho h(\bx),\varrho), 1),~ \text{s.t.}\!\sum_{n=1}^{N_0}\pi(\bx_n) \!\!=\!\! N_0\alpha.
% \end{align}

% Given the hardness score $h(\bx)$ for a sample $\bx$, we set the minimum sampling rate $\varrho$ and tune a linear scaling parameter $\rho$ to meet the average subsampling rate $\alpha$

% \az{Given the hardness score $h(\bx)$ for a sample $\bx$, we set the minimum sampling rate $\varrho$ and tune a linear scaling parameter $\rho$ to meet the average subsampling rate $\alpha$
% \begin{align}\hspace{-5pt}
%     \pi(\bx) \!\!=\!\! \mathrm{max}(\mathrm{min}(\rho h(\bx),\varrho), 1),~ \text{s.t.}\!\sum_{n=1}^{N_0}\pi(\bx_n) \!\!=\!\! N_0\alpha.
% \end{align}
% }

% Let $\pi_\calD(\bx)$ be the subsampling rate for $\bx$ by using model-agnostic hardness function $h(\cdot):=f(\cdot;\calD)$. And let $\pi_\phi(\bx)$ be the subsampling rate for $\bx$ by using pilot model $h(\cdot):=f(\cdot;\phi)$. 

We propose three simple yet effective heuristic strategies to combine these two methods and get the final sampling rate: \textit{maximum}, \textit{mean}, and \textit{product}.
\begin{align}
\label{eq:combination}
    \pi_{\mathrm{max}}(\bx) &= \rho_\mathrm{max}\mathrm{max}\big(\pi_\calD(\bx), \pi_{\phi}(\bx)\big);\\
    \pi_{\mathrm{mean}}(\bx) &= \frac{(\pi_\calD(\bx)+ \pi_{\phi}(\bx))}{2};\nonumber\\
    \pi_\mathrm{prod}(\bx)&=\mathrm{min}\Big(\mathrm{max}\big(\rho_\mathrm{prod}\pi_\calD(\bx)\pi_{\phi}(\bx)\big),~\varrho_\mathrm{prod}),1\Big).\nonumber
\end{align}
Note $\varrho_\mathrm{prod}$ is an extra hyperparameter when applying product combination. And $(\rho_\mathrm{max},\rho_\mathrm{prod})$ are tuned to normalize the average sample rate to $\alpha$.

% control the overall subsampling rate. 

% \jk{what's the correlation between equation ~\ref{eq:combination} and ~\ref{eq:pi_x}? if only $\pi_\calD(\bx)$ is calculated by equation ~\ref{eq:pi_x}, just use $\pi_\calD(\bx)$. Another concern of notation: one is $\calD$, one is $\phi$. how about a for agnostic and b for based.}

% With a reliable sample rate, we use Eqn.~(\ref{eq:logits_correction}) as the training objective to guarantee our result is well-calibrated. 
After the subsampling rate combination, each $\bx$ will be sampled with probability in Eqn.~(\ref{eq:combination}). All the sampled instances will follow the normal training protocol to optimize the training objective as shown in Eqn. ~(\ref{eq:logits_correction}) which guarantees the final result to be well-calibrated.

\vspace{0em}
\subsection{Theoretical bottlenecks of model-based subsampling}
\label{sec:bottleneck}
\ed{
% Above, we propose a principled approach by studying graph structures within data to figure out the sample's hardness without referring to a pilot model.
{Above, we propose a principled approach to determine the sample's hardness by analyzing graph structures within the data without referring to a pilot model.}
We advocate this model-agnostic sampling method due to the dilemma of unavoidable model misspecification in the online-offline discrepancy setup.
% A correctly specified pilot model is the critical assumption to derive a theoretically efficient data subsampling method.
The critical assumption of a correctly specified pilot model is needed to derive a theoretically efficient data subsampling method.
Under pilot model misspecifications, it is even non-trivial to guarantee model consistency (the subsampled estimator has the same limit as the full data estimator) for the log odds correction estimator (Eqn.~\ref{eq:logits_correction})~\citep{fithian2014local,han2020local, shen2021surprise}. Therefore, pilot model misspecification is a realistic but underexplored problem in statistics.
}

% \jk{do we have to talk about logits shift and calibration issues?}
\section{Experiments}
\label{sec:exp}\vspace{0em}
% Here we train models with data sampled via different subsample methods and compare their performance. First, we show that effective resistance fails to assign high subsample rates to hard negatives. Second, we show that pilot misspecification will lead to discrepancies in model performance. Third, we report the model performances over two datasets to demonstrate the superiority of our method. Finally, we conduct extensive ablation studies to investigate the effectiveness of the proposed components.
We compare various subsampling methods on downstream target model performance. First, for model-based sampling, we show that pilot misspecification will lead to discrepancies in model performance. Second, we report empirical results over two datasets to demonstrate the superiority of our model agnostic subsampling method. Third, we conduct extensive ablation studies to investigate the effectiveness of model-agnostic hardness score, score propagation, and the benefit of ensembling model-agnostic and model-based methods. Finally, we discuss effective resistance and its relationship to negative sampling.
\vspace{0em}

\subsection{Datasets and data pre-processing}\vspace{-0.0em}
We demonstrate our empirical results using KuaiRec~\citep{gao2022kuairec} and Microsoft News Dataset (MIND)~\citep{wu2020mind}. The statistics of the pre-processed datasets are shown in Table~\ref{tab:data-stats}. For both datasets, we use 80\% of the data for training, 10\% for validation, and 10\% for testing. We report all experimental results for 8 runs with random initializations on one random data split. We set the average subsampling rate $\alpha=0.2$ on training data for both datasets.

\paragraph{KuaiRec} KuaiRec is a recommendation dataset collected from the video-sharing mobile app ``Kuaishou". The dataset is generally a sparse user-item interaction matrix with a fully-observed small submatrix. We crop the fully-observed submatrix and only consider the rest entries in the sparse matrix since those data are collected under natural settings. We use the label ``watch\_ratio'', which represents the total duration of a user watching a video divided by the video duration. 
Since a user may watch a video multiple times, the watch ratio can be larger than 1. In the experiment, we consider a user like a video (positive) if the ``watch\_ratio'' is larger than 3.

\begin{table}[t]
    \centering
    \begin{tabular}{c|ccccc}\hline
        &\#users & \#items & \#instances &\#pos.:\#neg.\\\hline% &\#isolated edges in training graph\\\hline
        KuaiRec &  5,765 & 10,679 &11,959,745 & 1:35\\% & 291,484\\
        MIND & 94,057&22,771&8,236,715&1:25\\\hline% & 415,395\\\hline
    \end{tabular}\vspace{0em}
    \caption{Dataset statistics.}
    \label{tab:data-stats}
    \vspace{-2em}
\end{table}
\paragraph{MIND} MIND is a large-scale news recommendation dataset with binary labels indicating users' impressions of recommended news. We do not use the content data in each news corpus during the experiment. The MIND dataset does not require extra pre-processing.

\begin{figure}[t]
    \centering
    \includegraphics[width=0.48\textwidth]{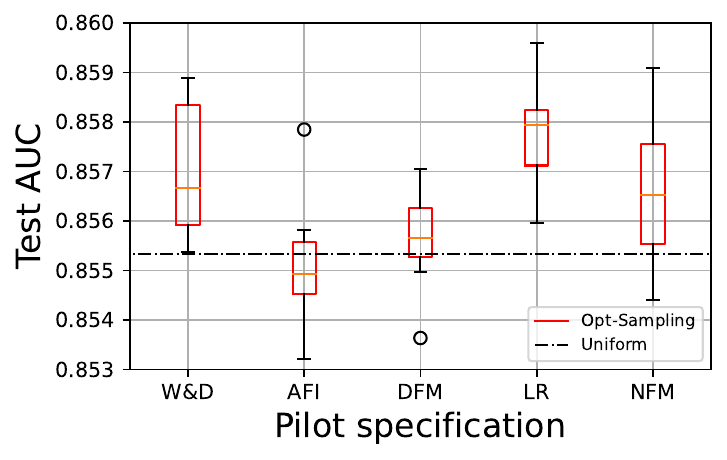}
    \vspace{-1em}
    \caption{Pilot misspecifications lead to inconsistent model performance in Opt-Sampling.}
    \label{fig:pilot-misspec}
    \vspace{-1em}
\end{figure}
\subsection{Baselines and model selections}

\paragraph{Baseline subsampling methods.} 
We consider two baselines -- the first baseline is the model-agnostic uniform negative sampling; the second is a model-based near-optimal sampling method (Opt-Sampling)~\citep{wang2021nonuniform}, which relies on the prediction scores of a pilot model as the hardness score to calculate sample rates.

\ed{We study the scenario where data subsampling rates are pre-computed and do not change during model training. This scenario is aligned with industrial applications such as recommendation systems, where negative data are subsampled before saving in distributed storage to get rid of huge saving costs. Our method contrasts previous studies to re-compute and subsample data on the fly. For example, SRNS~\cite{ding2020simplify} quantifies sample hardness by estimating the variance of the model prediction along training epochs, which is dynamically updated each time model sees the sample.
IRGAN~\citep{wang2017irgan} uses a GAN model where the generator draws negative samples from the negative pool and mixes them with positive samples. A discriminator classifies data and guides the generator to learn to sample hard negatives. These methods are not comparable to our work. To the best of our knowledge, Opt-Sampling~\citep{wang2021nonuniform} is the only representative method that allows for a fair comparison in the specific scenario considered by this work.}

\paragraph{Model architecture.} We choose the wide and deep model (W\&D)~\citep{cheng2016wide} as our training target model to validate the effectiveness of our model-agnostic method. For model-based subsampling methods, we need to pre-train a pilot model to estimate the sample hardness. To test the effect of pilot misspecification, we consider five types of pilot models: (1) W\&D model \citep{cheng2016wide}; (2) Linear logistic regression (LR); (3) Automatic feature interaction selection model (AFI)~\citep{liu2020autofis}; (4) Neural factorization machines (NFM)~\citep{he2017neural}; (5) Deep factorization machine (DFM)~\citep{guo2017deepfm}. In the rest of the experiments, unless otherwise specified, W\&D is used as the pilot model since it shares the same architecture as the target model (consistent pilot). All pilot models are trained using 10\% of the training data.

% \begin{table*}[t]
%     \centering
%     \begin{tabular}{c|cc|cc|cc|cc}\hline
%          &  Full &Uniform &Opt-Samp. &+Prop. &MA-EC(ours) &+Prop. &Comb.(Max.) & +Prop. \\\hline
%        & 0.862248 & 0.855330 & 0.857039&0.858348&0.857006&0.858192 &0.857747&0.858679\\
%            KuaiRec &$\pm$&$\pm$&$\pm$&$\pm$&$\pm$&$\pm$&$\pm$&$\pm$ \\
%             &0.002071&0.001070 & 0.001323&0.001388&0.000912 &0.001133 & 0.001209 &0.001067\\
            % &&&&&&&\\
%      &0.736120 &0.728429 & 0.728572 &0.727883& 0.729198 &0.729430& 0.729369 & 0.739628\\ 
%        MIND &$\pm$&$\pm$&$\pm$&$\pm$&$\pm$&$\pm$&$\pm$&$\pm$ \\
%             & 0.001523 & 0.000631 & 0.000652 &0.000404& 0.000659 &0.000763& 0.000386 & 0.000500\\\hline
%     \end{tabular}
%     \caption{Results on KuaiRec and MIND datasets}
%     \label{tab:all-results}
% \end{table*}

\subsection{Pilot misspecification}
\ed{Model-based subsampling methods can be sensitive to a misspecified pilot model. We fix the target model architecture and tune the pilot model architectures to study their effect on the KuaiRec dataset. Pilot models affect target model performance only through their generated samples. Figure~\ref{fig:pilot-misspec} demonstrates the target model performance with different pilot model specifications. Target models' test AUC varies from 0.8557 (AFI) to 0.8577 (LR) when changing pilots. The AUC difference is significant since the standard deviation is around 0.001, demonstrating a potential loss in large-scale recommendation systems processing millions of data points daily. This result consolidates the negative effect of pilot misspecification, which justifies using model-agnostic subsampling approaches.

% Model-based subsampling may rely on a correctly specified pilot model. We study the case on KuaiRec dataset when the pilot model is misspecified by using the same model-based subsampling approach while changing pilot model architectures. Figure~\ref{fig:pilot-misspec} reports the target model performance. We can observe that the target model's AUC obtained from different pilots varies from 0.8557 (AFI) to 0.8577 (LR). Since the standard deviation is around 0.001, the AUC difference is significant, demonstrating a potential loss in large-scale recommendation systems processing millions of data points on a daily basis. This result consolidates the effect of pilot misspecification, which justifies using model-agnostic subsampling approaches.

In Figure~\ref{fig:pilot-misspec}, there is a big difference in target models' test performance among misspecified pilot models. We further study the quality of pilot models w.r.t. their predictive performance. Table~\ref{tab:lr-overfit} demonstrates pilot models' training and testing AUC. Deep pilot models may overfit the training data since we use 1/10 data to pre-train pilot models. LR is less vulnerable to overfitting and achieves the highest test AUC. There is a caveat to using a misspecified pilot model, e.g. AFI in our case since the outcome can be worse than uniform sampling even when AFI can make reasonable predictions. Using a consistent pilot model (W\&D) is a safer choice: even when its predictive performance is worse than a misspecified NFM pilot model, the target model can benefit from pilot model consistency and achieve better testing AUC.

Note that W\&D is a strong architecture on the KuaiRec dataset. Table~\ref{tab:lr-target} shows that when using LR as the target model and trained on the same amount of data, its performance is worse than the one using W\&D as the target model: W\&D achieves 0.8553 in testing AUC with uniform sampling while LR gets 0.8474. Table~\ref{tab:lr-target} demonstrates both model-based sampling with a consistent pilot (Opt.Samp.) and model-agnostic sampling (MA-EC) can improve model performance. MA-EC consistently helps, disregarding the model specification. 
}

\begin{table}[t]
    \centering
    % \small
    \begin{tabular}{l|ccccc}\hline
         &  W\&D &AFI &DFM &LR &NFM\\\hline
         Pilot Train AUC$\uparrow$ & 0.9917& 0.9883 & 0.9922 & 0.9487 & 0.9999\\
         Pilot Test AUC$\uparrow$ & 0.8023 & 0.7919 & 0.7968 &0.8192& 0.8082\\\hline
    \end{tabular}
    % \vspace{1em}
    \caption{LR as a pilot model gives best pilot test AUC, thus yielding better subsampling performance}
    \label{tab:lr-overfit}
    \vspace{-1em}
\end{table}

\begin{table}[t]
    \centering

    \begin{tabular}{l|ccc}\hline
        &  Uniform &Opt-Samp. &MA-EC\\\hline
        Test AUC$\uparrow$ & 0.8474$\pm$0.0012& 0.8511$\pm$0.0009 &0.8519$\pm$0.0005 \\\hline
    \end{tabular}
    % \vspace{1em}
    \caption{LR as the target model: Model specification that is the best for a pilot model does not indicate it is the best for the target model.}
    \label{tab:lr-target}
    \vspace{-1em}
\end{table}

\begin{figure*}[t]
    \centering
    \includegraphics[width=\textwidth]{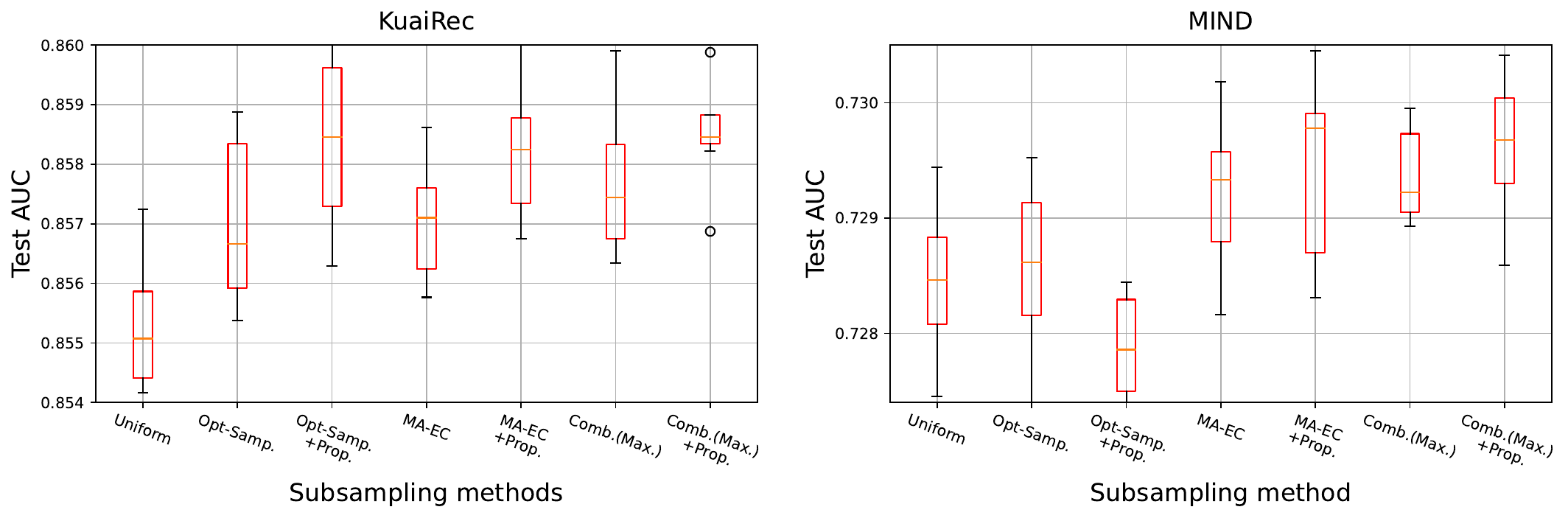}
    \vspace{0em}
    \caption{Experimental results on KuaiRec and MIND datasets. }
    \label{fig:all-results}
    \vspace{0em}
\end{figure*}
% \ed{Since we use 1/10 data to pre-train pilot models, deep pilot models may overfit the training data. LR is less vulnerable to overfitting and achieves the highest test AUC. Table~\ref{tab:lr-overfit} shows that LR works better simply because it does not suffer from overfitting. A good pilot model specification may not be a good target model specification. Table~\ref{tab:lr-target} shows that when using LR as the target model, the performance is worse than the one using W\&D as the target model. Another observation from Table~\ref{tab:lr-target} is that MA-EC consistently helps improve model performance, disregarding the model specification.}

\begin{figure}[t]
    \centering
    \includegraphics[width=0.48\textwidth]{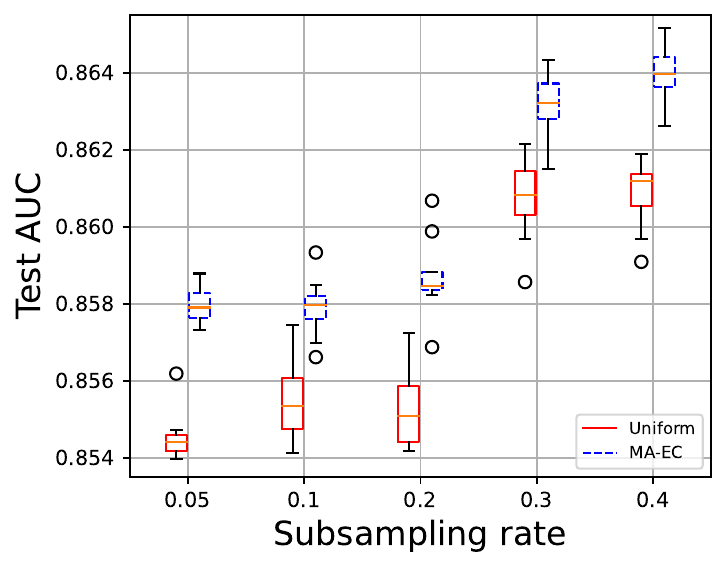}
    \vspace{0em}
    \vspace{-1em}
    \caption{Test AUC with different subsampling rates. MA-EC consistently leads to better model performance over different subsampling rates.}
    \label{fig:subsample-rate}
    \vspace{0em}
    \vspace{-1em}
\end{figure}

\subsection{Experiment results on two datasets}
We train the target model with different data sampling strategies and evaluate the model performance based on the area under the receiver operating characteristic curve (AUC). The results of all training configurations are presented in Figure~\ref{fig:all-results}. The comparison between the methods shows that the MA-EC strategy consistently outperforms the uniform sampling baseline in both datasets. Furthermore, when applied to the KuaiRec dataset, the MA-EC strategy achieved comparable performance to the Optimal Sampling method. However, in the MIND dataset, the Optimal Sampling approach did not improve upon the uniform sampling baseline and was outperformed by the MA-EC strategy. These results demonstrate the effectiveness of the MA-EC method in achieving improved performance in different datasets.

Additionally, the results were improved by incorporating a smoothing technique for hardness estimation through propagation. The performance of combining the model-agnostic and model-based methods using the maximum strategy was also reported. It was observed that the ensemble approach outperformed each individual method in both datasets. The smoothed scores from Opt-Sampling and MA-EC were also combined using the ensemble approach, resulting in the best performance as shown in the last column of the results. These findings highlight the benefits of combining multiple techniques to achieve even better performance in the target model training process. The use of smoothing and ensembling strategies demonstrates the potential for further improvements in data sampling methods and the effectiveness of combining multiple approaches to achieve better results.

\subsection{Ablation studies}
We use the KuaiRec dataset to conduct extensive ablation studies. First, we study if our method consistently outperforms baselines under different subsampling rates. Second, we show that MA-EC and Opt-Sampling can estimate sample hardness from different perspectives by designing a variable control experiment. And we investigate the effectiveness of different ensemble strategies. Third, we show how smoothing scores help further improve model performance.

\paragraph{Subsampling rate.} We compare uniform sampling with the best of our methods by ensembling smoothed scores from Opt-Sampling and MA-EC. Figure~\ref{fig:subsample-rate} demonstrates our methods consistently outperform uniform sampling under different subsampling rates.

\begin{table}[t]
% \small
    \centering
    \begin{tabular}{cc}\hline
       \multicolumn{2}{c}{ $\pi_\phi(\bx)$ as major score}\\
        no flip & flip $\pi_\phi(\bx)<0.2$ \& $\pi_\calD(\bx)>0.8$\\\hline
        0.8570 $\pm$ 0.0013 &0.8576 $\pm$ 0.0007\\\hline\\
        \multicolumn{2}{c}{ $\pi_\calD(\bx)$ as major score}\\
        no flip &flip $\pi_\calD(\bx)<0.2$ \& $\pi_\phi(\bx)>0.8$ \\\hline
        0.8570 $\pm$ 0.0009& 0.8574 $\pm$ 0.0011\\\hline
    \end{tabular}
    % \vspace{1em}
    \caption{Control experiment. Some negative samples have low subsampling rates in one method but have large subsampling rates in the other. Flipping those subsampling rates from small to large improves model performance.}
    \label{tab:control}
    \vspace{0em}
\end{table}

\begin{figure*}[t]
    \centering
    \begin{tabular}{c}
    \includegraphics[width=\textwidth]{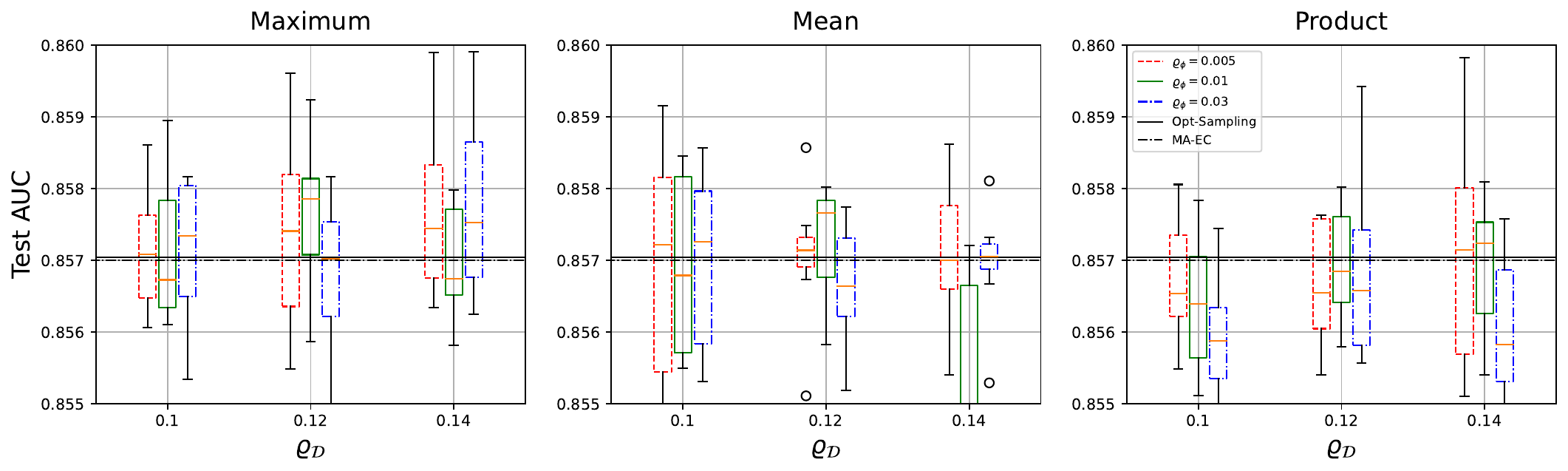}\vspace{0pt}\\
    (a) Model performance of maximum, mean, and product ensembles.\vspace{0pt}\\\\
    \includegraphics[width=\textwidth]{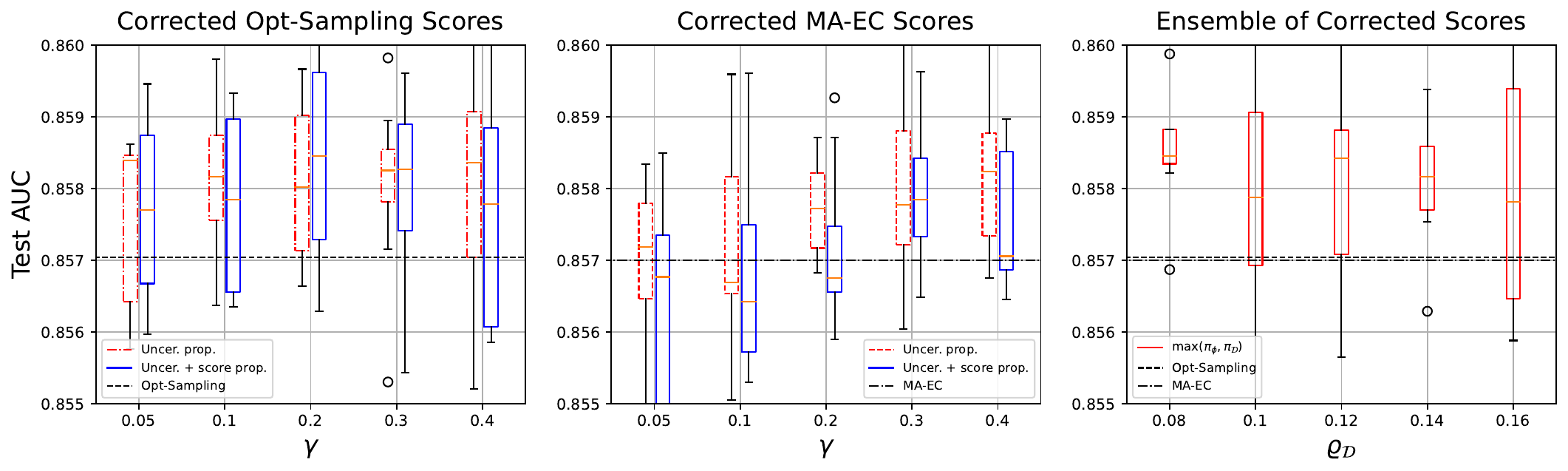}\vspace{0pt}\\
    (b) Model performance of corrected scores in Opt-Sampling, MA-EC, and their ensemble.
    \end{tabular}
    \vspace{0em}
    \caption{Abaltion studies on ensemble strategies and score correction.}
    \label{fig:ablation-comb-prob}
    \vspace{0em}
\end{figure*}

\paragraph{Ensemble strategies.}
To investigate whether hardness scores from MA-EC and Opt-Sampling complement each other, we design the following control experiment: (1) we assign instances with subsampling rates from each method; (2) for instances that have inconsistent subsampling rates between two methods, we flip their subsampling rates into the other method. For example, we assign instances that have $\pi_\calD(\bx)<0.2$ and $\pi_\phi(\bx)>0.8$ with the subsampling rate $\pi_\phi(\bx)$, and the rest instances with $\pi_\calD(\bx)$. Table~\ref{tab:control} shows the results of the control experiments, where we achieve better model performance by assigning most of the sample with one set of scores and flipping part of the sample scores. This verifies that some hard negative instances might be overlooked by one method and can be discovered by the other.

The control experiment justifies ensembling MA-EC and Opt-Sampling. We experiment with ensemble strategies (maximum, mean, and product). Figure~\ref{fig:ablation-comb-prob}(a) shows the box plot of the three ensemble methods. For each method, we present nine configurations of the hyperparameters ($\varrho_\calD, \varrho_\phi$), where $\varrho_\calD\in\{0.1,0.12,0.14\}$ and $\varrho_\phi\in\{0.005, 0.01, 0.03\}$. For product strategy, we set hyperparameter $\varrho_\mathrm{prod}=0.005$ for all experiments. We observe that the maximum strategy consistently gives comparable or better results than Opt-Sampling and MA-EC. While in mean and product strategies, we do not observe significant improvement. For the product strategy, the model performance even deteriorates. MA-EC needs to compute effective conductance to calculate subsampling rates. Effective conductance computation is not our bottleneck since it can be reused once computed. MA-EC is model-agnostic, and thus can support training with different target models.

\paragraph{Score correction.}
We investigate the effectiveness of correcting the hardness scores via graph propagation. We are interested in applying score correction in Opt-Sampling and MA-EC and their ensemble. Additionally, in applying both score correction and score ensemble, we can try either \textit{ensembling corrected scores} or \textit{correcting ensemble scores}. As we find the latter consistently results in worse performance, we only present the result of the former in this work. Figure~\ref{fig:ablation-comb-prob}(b) reports the model performance of our ablation study. For the experiments of correcting scores estimated from Opt-Sampling and MA-EC, we explore the propagation coefficient $\gamma\in\{0.05,0.1,0.2,0.3,0.4\}$. {For each coefficient, we iterate to smooth the scores until convergence}. Uncertainty propagation significantly improves model performance for both subsampling approaches. In score propagation, which runs on scores corrected by uncertainty propagation, model performance slightly improves in Opt-Sampling and worsens in MA-EC. In the ensemble of corrected scores, we combine the hardness scores from the best configurations of both methods via maximum strategy. The reported result demonstrates that the ensemble strategy improves model performance over not only the original scores but also the corrected scores.

\begin{figure}
    \centering
    \includegraphics[width=0.48\textwidth]{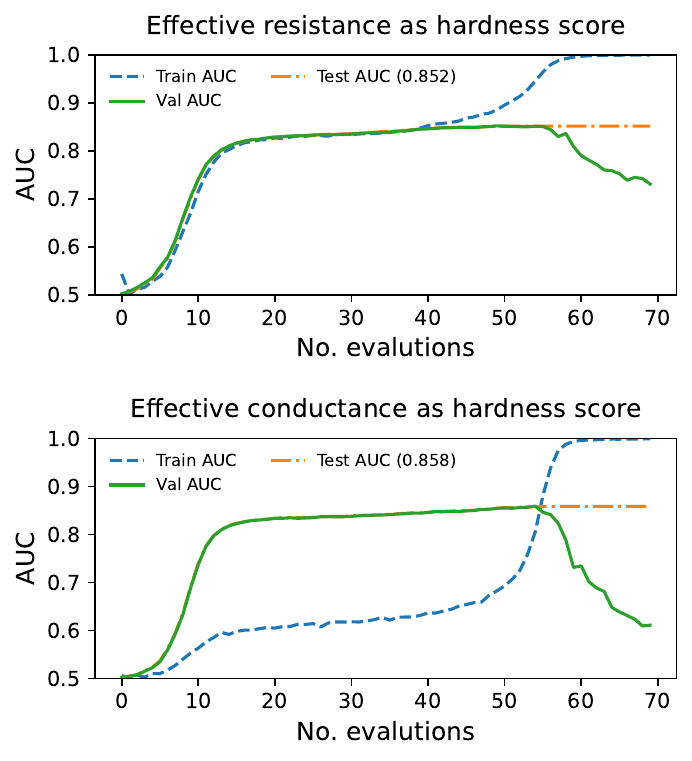}
    \caption{Comparison of a single target model run using effective conductance and effective resistance as hardness scores for negative sampling.}
    \label{fig:training-auc}
    \vspace{0em}
\end{figure}
\subsection{A note on effective resistance}
The effective resistance $R_\mathrm{eff}(u,v)=1/G_\mathrm{eff}(u,v)$ is often used for graph sparsification~\citep{spielman2008graph}. An edge with high effective resistance is considered important in maintaining graph topology. 
Since the definitions of edge importance using effective conductance and effective resistance run against each other, we demonstrate that defining edge importance with effective resistance is not applicable in our scenario.

We compare two model-agnostic (MA) subsampling methods with effective resistance (MA-ER) and effective conductance (MA-EC) as the hardness scores. The effective resistance is computed on the graph where all edges have unit resistances. We demonstrate that MA-ER fails to capture hard negative instances. On the KuaiRec dataset, MA-ER yields an average test AUC of 0.8535, which is worse than uniform sampling (0.8553). To unravel how MA-ER and MA-EC affect model training, we randomly select a run from each method to visualize the model training metrics. In Figure \ref{fig:training-auc}, when using MA-ER, training AUC remains the same as testing AUC before convergence. Besides, the model converges earlier. In sharp contrast, in MA-EC, there is a huge gap between training AUC and testing AUC. The gap shows that training instances are overall harder than those in the test set. This verifies that MA-EC discovers hard negatives while MA-ER does not.
% \jk{maybe introduce some computational cost here. i.e. running time comparison. and emphasize that some steps in our proposed approach are one-time running. since our are model-agonistic, can be support different model training.}
\section{Related Works}
\label{sec:related}

\paragraph{Negative sampling.}
Hard negative sampling is pervasively used in recommendation systems. PinSage~\citep{ying2018graph} shows using curriculum learning with negative sampling is effective, implying hard negatives are helpful in the late stage of training. Metasearch offline datastream combines documents with in-batch mismatched queries with the highest similarity scores as hard negative samples~\citep{huang2020embedding}. Hard negative sampling is also justified in theory. \citet{fithian2014local} uses a pilot model to select examples whose responses were rare given their features preferentially.~\citet{han2020local} explores local uncertainty sampling for multi-class classification.~\citep{wang2020logistic} proves that for generalized linear logistic regression models, the optimal negative sampling rate is proportional to the pilot prediction value.
There are also approaches~\citep{chen2017sampling, wu2019noise} that utilize item popularity to estimate the importance of negative instances.

\paragraph{Graph sparsification.}
Graph sparsification tries to drop nodes and edges while preserving the graph structure. \citet{ghosh2008minimizing,spielman2008graph} studied edge sampling with effective resistance to preserve graph Laplacian. \citet{satuluri2011local} used the Jaccard similarity score to measure node similarity and pick the top-K associated edges for each node by their similarity. Other methods include sampling edges by counting the number of its associate triangles~\citep{hamann2016structure}, quadrangles~\citep{nocaj2014untangling}, using local graph information~\citep{le2021edge}, or using deep neural networks to coarsen the graph~\citep{cai2021graph}. Graph sparsification has been explored to preserve or improve downstream task performance~\citep{zheng2020robust,wan2021graph}. However, these method requires an end-to-end training framework and thus cannot be used for continuous deployment where subsampling rates are static. To the best of our knowledge, model-agnostic methods based on graph sparsification are under-explored in the literature.

\section{Conclusion}
\label{sec:conclusion}
We propose model-agnostic hard negative subsampling methods using the effective conductance on the user-item bipartite graph as the hardness score. We further exploit the graph structure by score propagation. Our model-agnostic complements model-based optimal sampling and provide a sustainable and consistent data subsampling solution to real-world recommendation systems \ed{with long-term impact. We discuss the cost upon deployment, its social impact, and future directions.}
%\jk{Our proposed sampling methods maybe benefit neighbor selection for learning graph embeddings. We leave it as future work.}
\ed{
\paragraph{Deployment cost} Deploying a model-agnostic sampling service requires a database with graph computing engines that support effective conductance lookup given pairs of user-item ids, which is computationally cheap in practice. Online serving will not be affected, so there is no change in serving latency. Offline models are trained only on sub-sampled data to enjoy reduced computational costs. It takes 150 and 80 seconds to estimate the sampling rate for KuaiRec and MIND datasets, respectively. The cost of running MA-EC sampling is negligible compared to model training.

\paragraph{Social impact} With less data, one can train models with less GPU time and use less data storage. Given the quantity of data and the requirement to iterate the model periodically in the industry, the data subsampling method can significantly reduce the carbon footprint.

\paragraph{Future work} Our proposed sampling methods can be used to select better neighbors for learning graph embeddings. We can further improve MA-EC by (i) exploiting context information to identify hard negative samples. (ii) utilizing negative edges when calculating effective conductance. We leave those as future work.}

\bibliographystyle{ACM-Reference-Format}
\bibliography{references}

% \onecolumn
\appendix

\begin{table*}[t]
    \centering
    \begin{tabular}{l|ccccc}\hline
    &\multicolumn{5}{c}{KuaiRec}\\
         & \#epoch & batch size & learning rate & Optimizer & size of training set\\\hline
        pilot model training & 15 & 512 & 0.001 & Adam & 10\%\\
        target model training & 100 & 1024 & 0.0001 & Adam & 80\%\\\hline
    &\multicolumn{5}{c}{MIND}\\
         & \#epoch & batch size & learning rate & Optimizer & size of training set\\\hline
        pilot model training & 80 & 512 & 0.001 & Adam & 10\%\\
        target model training & 200 & 1024 & 0.0001 & Adam & 80\%\\\hline
    \end{tabular}
    \caption{Training details of the two datasets}
    \label{tab:detail-training}
\end{table*}
\begin{figure*}[t]
    \centering
    \includegraphics[width=0.6\textwidth]{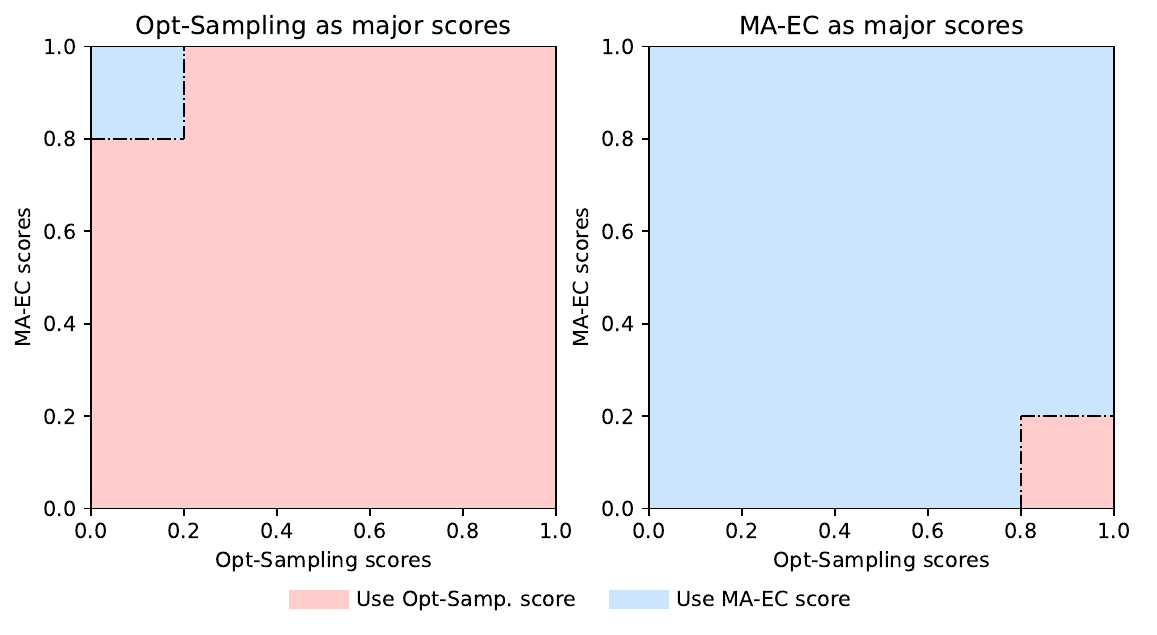}
    \caption{Illustration of the control experiment}
    \label{fig:control-demo}
\end{figure*}
\section{Implementation details}

\subsection{Dataset pre-processing}
Here we demonstrate the data pre-processing of the two datasets. The KuaiRec dataset contains a sparse interaction matrix and a dense interaction matrix, where the dense interaction submatrix of the sparse matrix. The sparse matrix contains 7,176 users and 10,728 items, while the dense matrix contains 1,411 users and 3,327 items. As mentioned, the dense matrix is not collected under the natural recommendation setting, and shaving off the submatrix from the whole matrix is also unrealistic. Specifically, we create a submatrix by removing all the user columns from the sparse matrix if they appear in the dense matrix. The origin MIND-small dataset contains a training set and a validation set. In our setting, we first merge the two sets and re-split them into training, validation, and testing set. For both datasets, user-item pairs might appear multiple times in the dataset under a different context. When constructing the user-item bipartite graph, we treat them as \textit{one} edge with a conductance of 1 if any of the pairs is positive and 0 otherwise. We assign all duplicate pairs with the same subsampling rate.

\subsection{Model specifications and hyperparameters}
For pilot models, we adopt the default configuration in TorchFM\footnote{See \href{https://github.com/rixwew/pytorch-fm}{https://github.com/rixwew/pytorch-fm}} except for D\&W. The training details of the pilot and the target models are shown in table~\ref{tab:detail-training}. To compute effective conductance over the graph, we run random walk simulation until the update of commute time is smaller than 0.1. In edge propagation, the best results are reported with $\gamma=0.2$ and $\gamma=0.4$ for Opt-Sampling and MA-EC, respectively, in the KuaiRec dataset. And $\gamma=0.05$ for both methods in the MIND dataset.
\subsection{Control experiment}
We illustrate the control experiment in Figure~\ref{fig:control-demo}. In each setting, one set of the scores is used as the major sampling scores, and only part of the instances will use the other ones. For example, when Opt-Sampling is used as major scores (red rectangle), instances whose scores lie in the upper left (blue rectangle) have inconsistent hardness over the two methods. And since MA-EC considers them hard negative samples, we ``flip'' the subsampling score of those instances from Opt-Sampling to MA-EC. 
% put the plot in the our doc here
\begin{table}[h]
    \centering
    \begin{tabular}{l|ccc}\hline
        &  Uniform &Opt-Samp. &MA-EC\\\hline
        % ACC$\uparrow$&0.9743$\pm$0.0004 &0.9747$\pm$0.0003 &0.9747$\pm$0.0004\\
        ACE$\downarrow$&0.0070$\pm$0.0005 &0.0073$\pm$0.0010&0.0071$\pm$0.0014\\
 NDCG@5$\uparrow$ & 0.5209$\pm$0.0034 & 0.5342$\pm$0.0021 & 0.5390$\pm$0.0034\\
   NDCG@10$\uparrow$ & 0.5223$\pm$0.0019 & 0.5375$\pm$0.0018 & 0.5403$\pm$0.0023\\
   NDCG@30$\uparrow$ & 0.5221$\pm$0.0022 & 0.5364$\pm$0.0026 & 0.5397$\pm$0.0030\\\hline
    \end{tabular}
    \caption{Model performance on offline metrics}
    \label{tab:offline}
\end{table}
\subsection{Computing resources}
For all the experiments, we train the models on NVIDIA A100 and NVIDIA V100 GPUs. For estimating the effective conductance on the graph, we utilize 16 CPU cores and 40 gigabytes of memory to run the simulations for both datasets. 
molecule generation tasks, the inference time of each model is measured on 1 TITAN RTX GPU and 20 CPU cores.

\begin{figure*}[h]
    \centering
    \begin{tabular}{c} 
    \includegraphics[width=\textwidth]{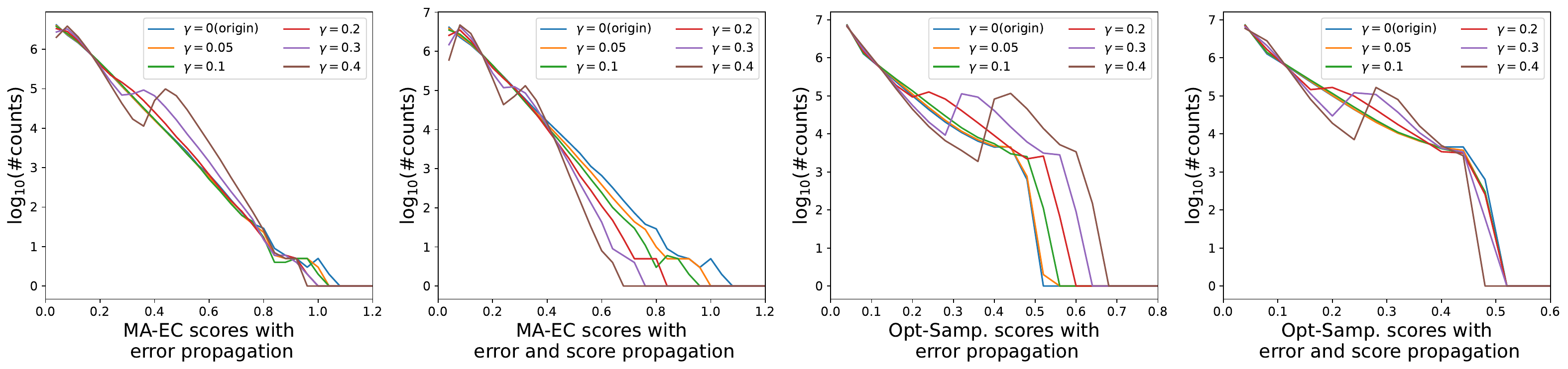} \\
    KuaiRec\\
    \includegraphics[width=\textwidth]{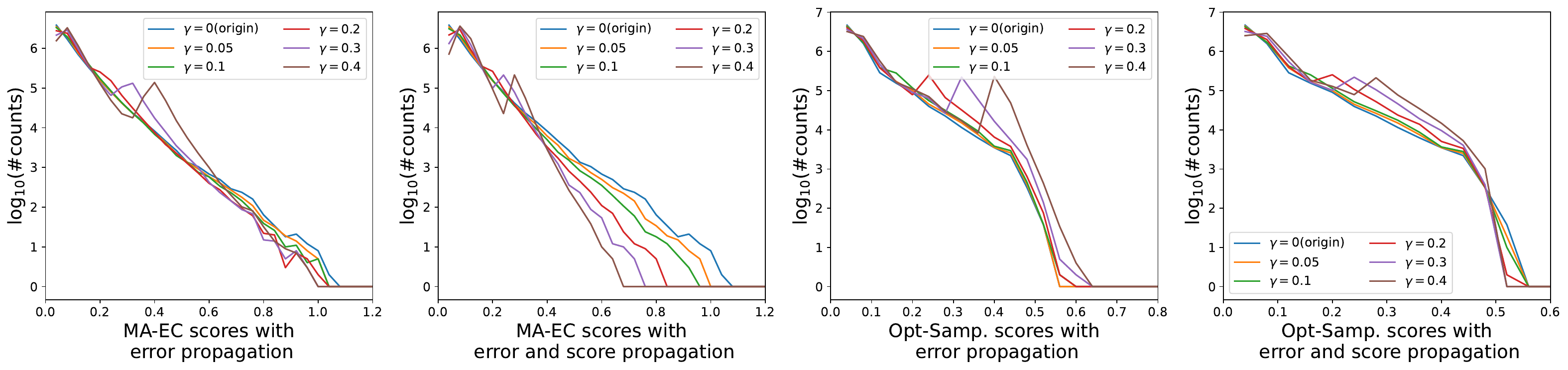} \\
    MIND \\
    \end{tabular}
    \caption{Distribution of propagated scores.}
    \label{fig:score_hist}
\end{figure*}
\section{Additional results}
\subsection{Model performance on prediction tasks}
We further investigate the performance gain of MA-EC in other offline metrics. Specifically, we consider Normalized Discounted Cumulative Gain (NDCG) scores and adaptive calibration error (ACE). We consider uniform sampling, Opt-Sampling, and MA-EC on the KuaiRec dataset in Table~\ref{tab:offline}. We see that MA-EC and Opt-Sampling outperform uniform in all metrics by around 0.015 in terms of NDCG, illustrating the benefit of non-uniform sampling. As ACE measures whether the probabilities predicted by the classifier are calibrated~\citep{nixon2019measuring}, we can see that our model is well-calibrated even though the subsampled training set has a different population than the actual training set.
% \textbf{$\bullet$ How much gain we have in other reco offline metric?}\\
% % We report NDCG scores for uniform sampling, Opt-Sampling, and MA-EC on the KuaiRec dataset in the table below. We see that MA-EC and Opt-Sampling outperform uniform in all metrics by around 0.015, illustrating the benefit of non-uniform sampling.
% \begin{table}[h]
%     \centering
%     \small
%     \begin{tabular}{l|ccc}\hline
% & Uniform & Opt-Samp. & MA-EC \\\hline
%    NDCG@5$\uparrow$ & 0.5209$\pm$0.0034 & 0.5342$\pm$0.0021 & 0.5390$\pm$0.0034\\
%    NDCG@10$\uparrow$ & 0.5223$\pm$0.0019 & 0.5375$\pm$0.0018 & 0.5403$\pm$0.0023\\
%    NDCG@30$\uparrow$ & 0.5221$\pm$0.0022 & 0.5364$\pm$0.0026 & 0.5397$\pm$0.0030\\\hline
%         %  & 
%         % NDCG@5$\uparrow$ & NDCG@10$\uparrow$ & NDCG@30$\uparrow$\\\hline
%         % Uniform & 0.5209$\pm$0.0034& 0.5223$\pm$0.0019 &0.5221$\pm$0.0022\\
%         % Opt-Samp.&0.5342$\pm$0.0021&0.5375$\pm$0.0018&0.5364$\pm$0.0026\\
%         % MA-EC &0.5390$\pm$0.0034&0.5403$\pm$0.0023&0.5397$\pm$0.0030\\\hline
%     \end{tabular}
%     \label{tab:my_label}
% \end{table}

\subsection{Smoothness of propagated scores}
For both datasets, we visualize the distributions of the corrected hardness scores for both MA-EC and Opt-Sampling methods. Specifically, we divide the instance scores into 50 bins and count the number of instances (both positives and negatives) for each bin. We generate the histograms and visualize them in Figure~\ref{fig:score_hist}. 

\end{document}